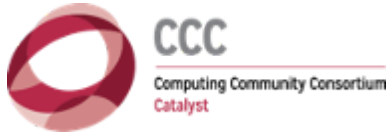

# Post Quantum Cryptography: Readiness Challenges and the Approaching Storm

*A Computing Community Consortium (CCC) Quadrennial Paper*

*Matt Campagna (Amazon), Brian LaMacchia (Microsoft Research), and David Ott (VMware Research)*

## Introduction

While advances in quantum computing promise new opportunities for scientific advancement (e.g., material science and machine learning), many people are not aware that they also threaten the widely deployed cryptographic algorithms that are the foundation of today's digital security and privacy. From mobile communications to online banking to personal data privacy, literally billions of Internet users rely on cryptography every day to ensure that private communications and data stay private. Indeed, the emergence and growth of the public Internet and electronic commerce was arguably enabled by the invention of *public-key cryptography*. The key advantage offered by public-key cryptography is that it allows two parties who have never communicated previously to nevertheless establish a secure, private, communication channel over a non-private network (e.g., the Internet). Public-key cryptography is also the technology that enables digital signatures which are widely used to protect software and application updates, online contracts, and electronic identity documents like Personal Identity Verification (PIV) credentials and e-passports.

Underlying modern cryptography are specific *algorithms* which describe how mathematical objects and operations can be used to generate cryptographic keys, perform encryption and decryption, and compute digital signatures. Today's algorithms were selected for standardization and widespread deployment because of their robust security properties which prevent an adversary from reverse engineering data and secrets, even with the most advanced computer hardware available today.

Nevertheless, the security of an individual cryptographic algorithm can weaken over time as advances in cryptographic analysis and computing technology empower an adversary in new ways as they look to attack an existing implementation or algorithmic standard. When we lose confidence in a cryptographic algorithm, we must identify a replacement algorithm and transition all current uses of it to a suitable replacement.

Recent advances in quantum computing signal that we are on the cusp of our next cryptographic algorithm transition, and this transition to *post-quantum cryptography* will be more complicated and impact many more systems and stakeholders, than any of the prior migrations. This transition represents a major disruption within the IT industry and will broadly impact nearly every domain of our digital lives, from global commerce to social media to government and more. Cryptographic algorithm transitions take time and involve an extensive coordination effort across many stakeholders who are involved in building and operating the world's compute infrastructure. By preparing now for the upcoming transition to these new algorithms, we can ensure a more orderly, less costly, and minimally disruptive changeover.

## The Implications of Quantum Computers

Since 1994, we have known that all of our widely deployed public-key cryptographic algorithms can be attacked efficiently with a large enough quantum computer, but whether a quantum computer could even be built was a purely theoretical question. While today's quantum computers are not big enough or stable enough to threaten our current cryptographic algorithms, they point the way to future devices that will. Furthermore, while a *cryptographically relevant* quantum computer may not be realized for a decade or longer, its future existence is a threat to the security of information we send and receive today due to the *capture now, exploit later* attack. Long-lived information assets that are encrypted with today's algorithmic standards may be recorded and stored by an adversary as parties exchange information over the network. The adversary can then exploit the data later when large-scale quantum computers are available. The threat of the *capture now, exploit later* means we need to be using quantum-resistant algorithms well in advance of the scaled quantum computers necessary to fully realize the attack.

Acknowledging the threat of quantum computers to existing cryptography, the US National Security Agency (NSA) published warnings of the need to transition to new quantum-resistant algorithms in 2015, and in 2017 the US National Institute of Standards and Technology (NIST) launched a standardization initiative to select quantum safe algorithms for future use by government and industry. Referred to as *post quantum cryptography*, the new algorithm proposals are in the third round of analysis and vetting. NIST is expected to announce the first algorithms to qualify for standardization within 18-24 months, with a Federal Information Processing Standard (FIPS) for these algorithms to follow within a year. In its progress report announcing Round 3 selections, NIST said there may also be a Round 4 for additional algorithms, so it is possible that multiple FIPS will be issued for PQC algorithms over the next five years.

## The Road Ahead

The publication of a FIPS for one or more post-quantum algorithms is not the end of the standardization process but just the completion of its first phase. We know from experience with prior algorithm migrations (which were all simpler by comparison) that it takes considerable time and effort for an entire industry to come together and update protocol standards and deploy them using the new algorithms. In practice, each protocol has its own standardization context and community and updating is a bespoke process. Industry experience with prior cryptographic algorithm migrations shows that even apparently simple replacements take significant time; for example, the industry migration from the SHA-1 hash function (deprecated in 2011) to its replacement SHA-2 is still ongoing. The post-quantum cryptography transition will mean a wide array of standards will need to be updated for applications impacting the daily lives of citizens including electronic payments, online banking, remote learning, mobile communications, streaming content and entertainment, satellite communications, emergency services and critical infrastructure, and more.  In many cases the process of migrating to post-quantum cryptography in each of these areas will involve changes to a chain of dependent standards plus end-user devices with activity occurring in multiple international standards organizations.  Thus, whatever work we can front-load now, independent of specific algorithm choices, will help us speed up the migration process.

Given the progress NIST is making on standardizing post-quantum cryptographic algorithms and the amount of standards work that will follow, now is the time for every government agency and standards-dependent organization to start planning for their transition to post-quantum cryptographic algorithms.  Much of the post-quantum transition work that needs to be done can start now, before we know which specific algorithms NIST will choose to standardize.  Agencies can start by understanding the lifecycles of the cryptographic standards relevant to their specific activities and what work those standards organizations are already doing to get ready for the post-quantum migration.  The most important step we can take now is ensuring relevant stakeholders across government agencies understand the post-quantum algorithm transition to come, the impact it will have on their existing systems, and support required for new post-quantum algorithms in every computing system architected and designed from this point forward.  ***Awareness is key – simply knowing to ask the question, "Is this system designed to accommodate the transition to post-quantum cryptography?" is crucial.***

In addition, agencies should start the engineering work now to prepare their systems for the coming migration, including running real-world trials and tests of their systems modified to use post-quantum cryptographic algorithms (using some of the NIST Round 3 algorithms as placeholders for the eventual standards).  This is an especially important activity for owners of critical infrastructure.  For projects that have planned deployment and activity dates beyond the expected completion of the first post-quantum cryptography FIPS in 2024, the ability to deploy with post-quantum algorithms should be a non-negotiable requirement.  For all these activities, the key feature to ensure in new systems and RFPs is *cryptographic agility.*  Cryptographic agility is the ability to reconfigure a system or application or protocol with a different cryptographic algorithm, including algorithms not yet defined at the time the system was initially constructed.  Taking steps now to ensure that future systems under development have sufficient cryptographic agility will avoid future pain.

This is also the time to **sponsor and incentivize new academic research** to help with the many challenges associated with cryptographic algorithm migrations and the building of cryptographically agile systems.  Note that the new algorithms under standards consideration by NIST for standardization will have significantly different parameters (key sizes, ciphertext sizes, signatures sizes) and computational requirements (memory, processing cycles, network latencies), as well as some new requirements (e.g., entropy, failure handling).  As such, research is needed to examine the implications for a wide variety of cryptography usage domains (e.g., secure communications, authentication, key management, web applications, mobile and IoT computing) and to understand performance challenges.

Automated tools are needed to inventory cryptography usage across compute infrastructure, to migrate algorithm implementations in scaled infrastructure, and to test the results of migration.

Cryptographic agility is an area often overlooked in computer systems research yet critically important to secure and efficient algorithm migration.  Research in new tools and frameworks for ensuring cryptographic agility that can be configured and controlled by policy are sorely needed.  Gaps are especially notable in agility frameworks for large-scale infrastructures (e.g., private data centers and public cloud) and distributed application architectures (e.g., Web services, distributed container schemes).  Agility frameworks must not only enable policy-driven configuration, they must also be protected from adversarial tampering and attack.  How agility creates new attack surfaces and how those attack surfaces will be protected is another key area of research.

We recommend encouraging new research in these areas through an array of existing government funding mechanisms, university involvement, and industry partnerships.

**Summary**

Quantum computers are in our future, even if we do not know today when they will achieve scale or become commonplace and accessible.  While they promise to enable vast new computational workloads that are not feasible with classical computing, they also represent an important threat to our widely deployed cryptography algorithms and standards.  Because the information we encrypt and transmit with today's public-key cryptography can be recorded by an adversary and stored for future exploitation using quantum computers, concern over the implications of quantum computing for today's cryptography has urgency in the present.  The sooner we can migrate to post-quantum cryptographic algorithms, the more quickly we can move to counter this future security threat impacting the global Internet as we know it.  This algorithm migration will be more involved and complicated than any we have faced in the past, so early awareness, education, and planning are key.  The more we can do today to prepare for this disruption, the easier our overall migration journey will be as an industry.

*This white paper is part of a series of papers compiled every four years by the CCC Council and members of the computing research community to inform policymakers, community members and the public on important research opportunities in areas of national priority. The topics chosen represent areas of pressing national need spanning various subdisciplines of the computing research field. The white papers attempt to portray a comprehensive picture of the computing research field detailing potential research directions, challenges and recommendations.*

*This material is based upon work supported by the National Science Foundation under Grant No. 1734706. Any opinions, findings, and conclusions or recommendations expressed in this material are those of the authors and do not necessarily reflect the views of the National Science Foundation.*

*For citation use: Campagna M., LaMacchia B., & Ott D. (2020) Post Quantum Cryptography: Readiness Challenges and the Approaching Storm.*
*https://cra.org/ccc/resources/ccc-led-whitepapers/#2020-quadrennial-papers*